\documentclass[11pt]{article}
\usepackage[utf8]{inputenc}
\usepackage{amsmath}
\usepackage{amsfonts}
\usepackage{mathrsfs}
\usepackage{mathtools}
\usepackage[dvipdfmx]{graphicx}
\usepackage[left=28mm,
right=28mm,
top=30mm,
bottom=30mm]{geometry}
\usepackage{here}
\usepackage{amsthm}
\usepackage{algorithmic}
\usepackage{algorithm}
\usepackage{url}
\usepackage{stmaryrd}
\usepackage[title]{appendix}
\usepackage{amssymb}
\usepackage[all]{xy}
\usepackage{amscd}
\usepackage{color}
\usepackage{extarrows}
\usepackage{multirow}
\usepackage{enumitem}
\usepackage[
        dvipdfmx,
        hypertexnames=false,
        hyperindex,
        bookmarks=false,
        colorlinks=true,
        linkcolor=magenta,
        citecolor=magenta,
        urlcolor=blue
]{hyperref}
\usepackage{bm}

\newtheorem{theorem}{Theorem}[subsection]

\theoremstyle{definition}

\newtheorem{remark}[theorem]{Remark}

\theoremstyle{plain}

\theoremstyle{plain}

\theoremstyle{plain}
\newtheorem{prop}{Proposition}
\theoremstyle{plain}

\theoremstyle{definition}

\numberwithin{equation}{section}

\makeatletter
\newcommand*{\xnRightarrow}[2][]{%
  \ext@arrow 0359\nRightarrowfill@{#1}{#2}%
}
\newcommand*{\nRightarrowfill@}{%
  \narrowfill@\Relbar\Relbar\Rightarrow\neq
}
\newcommand*{\narrowfill@}[5]{%
  $\m@th\thickmuskip0mu\medmuskip\thickmuskip\thinmuskip\thickmuskip
  \relax#5#1\mkern-7mu%
  \cleaders\hbox{$#5\mkern-2mu#2\mkern-2mu$}\hfill
  \mkern-5mu %
  #4%
  \mkern-5mu %
  \cleaders\hbox{$#5\mkern-2mu#2\mkern-2mu$}\hfill
  \mkern-7mu#3$%
}
\makeatother

\title{Explicit cost analysis of Toom-4 multiplication for\\
incomplete NTT in lattice-based cryptography}
\date{\today}
\author{
{\Large Sakura Oku and Momonari Kudo}\\[2mm]
Department of Information and Communication Engineering\\
Fukuoka Institute of Technology\\
3-30-1 Wajiro-higashi, Higashi-ku, Fukuoka 811-0295, Japan\\
\texttt{s24b2043@bene.fit.ac.jp, m-kudo@fit.ac.jp}
}
\begin{document}

\maketitle

\begin{abstract}
Polynomial multiplication is fundamental in lattice-based cryptography. 
While the Number Theoretic Transform (NTT) enables fast multiplication, it imposes constraints on the modulus of the coefficient field.
Hafiz et al.\ (2025) addressed this limitation by analyzing the incomplete NTT, which combines a truncated NTT with conventional multiplication methods.

In this work, we revisit Toom-4 multiplication in the context of incomplete NTT. 
Although Toom-4 is asymptotically faster than Karatsuba, its precise cost has not been expressed in a form compatible with the incomplete NTT framework.

We present a concrete Toom-4 implementation and derive explicit operation counts that separate additions/subtractions and multiplications over the coefficient field. 
Our analysis based on addition chains yields a simple cost model for incomplete NTT.

Using this model, we analyze hybrid strategies combining Toom-4, Karatsuba, and incomplete NTT.
We identify parameter ranges where Toom-4 is advantageous and validate the predicted behavior experimentally.
\end{abstract}

%=====================
\section{Introduction}
%=====================

In practical lattice-based cryptosystems such as Kyber (ML-KEM)~\cite{Kyber,FIPS203} and Dilithium (ML-DSA)~\cite{Dilithium,FIPS204}, computations are performed over residue rings of the form $R_q = \mathbb{F}_q[x]/\langle x^n+1\rangle$, where $q$ is an odd prime and $n$ is a $2$-power.
% The Number Theoretic Transform (NTT) provides fast polynomial multiplication in this setting, but requires the condition $2n \mid (q-1)$. 
The full, or complete, Number Theoretic Transform (NTT) provides fast polynomial multiplication in this setting, but requires the condition $2n\mid(q-1)$.
This restriction can limit the choice of parameters and affect other design goals such as security and efficiency.

To address this issue, Hafiz et al.~\cite{INTT} formulated and analyzed the incomplete NTT, which relaxes this constraint by combining a truncated NTT with conventional polynomial multiplication methods.
Their work provides a cost model showing that such hybrid strategies can improve performance when the full, or complete, NTT is not available.
In this context, it is natural to consider fast multiplication algorithms such as Karatsuba~\cite{Karatsuba} and Toom--Cook~\cite{Toom,Cook} within their incomplete NTT framework. 

While the complexity of Karatsuba is well understood, the precise cost of Toom-$4$ -- especially its constant factors -- has not been expressed in a form compatible with the cost model of incomplete NTT. 
Although prior work, such as Bodrato and Zanoni~\cite{BZ2007}, studied optimized evaluation and interpolation schemes for Toom--Cook multiplication, these analyses are not directly suitable for integration into the incomplete NTT framework of Hafiz et al.~\cite{INTT}.

% The main contribution of this paper is to bridge this gap.
% We present a concrete Toom-$4$ implementation and derive explicit operation counts that separately track additions and multiplications in the coefficient field.
% Our analysis is based on addition-chain constructions, which enable a systematic determination of constant factors and yield a simple cost model compatible with incomplete NTT.
% Our contribution is not a new multiplication algorithm, but a precise integration of Toom--Cook into the incomplete NTT framework through explicit constant-level analysis.
% To the best of our knowledge, this is the first explicit Toom-4 cost model that is fully compatible with the incomplete NTT framework, enabling direct integration into its operation-count analysis.

The main contribution of this paper is to bridge this gap.
% We present a concrete Toom-$4$ implementation and derive explicit operation counts that separately track additions/subtractions and multiplications in the coefficient field.
We present a concrete Toom-$4$ implementation and derive explicit operation counts separately tracking additions/subtractions and multiplications in the coefficient field.
Our analysis is based on addition-chain constructions, which enable a systematic determination of constant factors.
The resulting cost model can be directly integrated into the operation-count framework of incomplete NTT.
Our contribution is not a new multiplication algorithm, but a precise constant-level analysis of Toom-$4$ multiplication tailored to the incomplete NTT framework.
To the best of our knowledge, such an explicit Toom-$4$ cost model has not been previously integrated into the incomplete NTT framework.

Using this refined model, we analyze hybrid multiplication strategies combining Toom-$4$, Karatsuba, and incomplete NTT.
This allows us to identify parameter ranges where Toom-$4$ becomes advantageous and to determine the optimal recursion depth.
We also provide a Python implementation, together with verification scripts for the addition-chain counts, and experimentally confirm the predicted behavior for NTT-unfriendly prime moduli near the Dilithium modulus.
% We also implement the proposed hybrid method in Python and conduct preliminary experiments for several NTT-unfriendly prime moduli near the Dilithium modulus.
% The experimental results are consistent with the theoretical predictions.
The Toom-$4$/Karatsuba hybrid is most effective when the admissible NTT depth is small.
The implementation and the scripts for verifying the addition-chain optimality are available at~\cite{GitHub}.
% Our study was also motivated by recent implementation-oriented works on polynomial multiplication in lattice-based cryptography, such as~\cite{SCIS2025}.

% Using this refined model, we analyze hybrid multiplication strategies combining Toom-$4$, Karatsuba, and incomplete NTT.
% This allows us to identify parameter ranges where Toom-$4$ becomes advantageous and to determine the optimal recursion depth.
% We also provide a Python implementation, together with verification scripts for the addition-chain counts, and experimentally confirm that the Toom-$4$/Karatsuba hybrid is most effective when the admissible NTT depth is small.

%======================
\section{Preliminaries}
%======================

This section reviews incomplete NTT and Toom--Cook multiplication used throughout the paper.

%=====================
\subsection{Notation}
%=====================

Let $K$ be a field, and $K[x]$ the univariate polynomial ring over $K$.
For $\phi(x)\in K[x]$, we write $\langle \phi(x)\rangle$ for the principal ideal generated by $\phi(x)$.
For $f(x)\in K[x]$ and nonzero $\phi(x)\in K[x]$, the remainder of $f(x)$ modulo $\phi(x)$ is denoted by $f(x)\bmod \phi(x)$.
We identify an element of $K[x]/\langle \phi(x)\rangle$ with its canonical representative of degree less than $\deg (\phi)$.
For a prime power $q$, we denote by $\mathbb F_q$ the finite field of order $q$.
When $q$ is fixed, let $\mathsf A$ and $\mathsf M$ denote one addition/subtraction and one multiplication in $\mathbb F_q$, respectively.

In lattice-based cryptography, one often works over residue rings of the form $\mathbb F_q[x]/\langle \phi(x)\rangle$.
A typical choice is the $2n$-th cyclotomic polynomial $\phi(x)=x^n+1$ with $n$ a $2$-power, since such rings admit efficient arithmetic and fast polynomial multiplication.
This setting is used in Ring-LWE and Module-LWE based cryptosystems such as Kyber (ML-KEM)~\cite{Kyber,FIPS203} and Dilithium (ML-DSA)~\cite{Dilithium,FIPS204}.
Hence, fast multiplication in $R_q=\mathbb F_q[x]/\langle x^n+1\rangle$ is of central practical importance.

%==========================
\subsection{Incomplete Number Theoretic Transform}\label{subsec:IncompleteNTT}
%==========================

The {\it Number Theoretic Transform (NTT)} is a finite-field analogue of the Fast Fourier Transform (FFT)~\cite{FFT,NTT}; see also~\cite{NTTsurvey}.
It is used to realize fast polynomial multiplication over residue rings, with $O(n\log_2 n)$ arithmetic operations over $\mathbb{F}_q$ for input degree less than $n$.
%denotes the polynomial degree (or transform length).
Let $q$ be an odd prime, let $n=2^k$, and put $R_q=\mathbb{F}_q[x]/\langle x^n+1\rangle$.
The full, or complete, NTT over $R_q$ requires a primitive $2n$-th root of unity in $\mathbb{F}_q$, equivalently $2n\mid(q-1)$.
The {\it incomplete NTT}~\cite{INTT} relaxes this condition by using only depth $\ell$, where $0\le \ell\le k$ and $2^{\ell+1}\mid(q-1)$, and then combining the transform with another conventional polynomial multiplication method.
The case $\ell=k$ gives the full NTT, while $\ell=0$ gives no transform.
Moreover, incomplete NTT can be implemented non-recursively and in-place on the coefficient representation of $R_q$, see, e.g.,~\cite{FIPS203,INTT}.

Assume that $2^{\ell+1}\mid(q-1)$.
We choose and fix a primitive $2^{\ell+1}$-th root of unity $\zeta_{2^{\ell+1}}$ in $\mathbb{F}_q$.
Then $\zeta_{2^{\ell+1}}^{2j+1}$ ($0\leq j \leq 2^{\ell}-1$) are exactly the primitive $2^{\ell+1}$-th roots of unity, and we have $x^{n}+1 = \prod_{j=0}^{2^\ell-1}\left(x^{n/2^\ell}-\zeta_{2^{\ell+1}}^{2j+1}\right)$.
% In particular, for each pair $(j,j')$ with $0 \leq j < j' \leq 2^t-1$, the polynomials $x^{n/2^t} - \zeta_{2^{t+1}}^{2j+1}$ and $x^{n/2^t}-\zeta_{2^{t+1}}^{2j'+1}$ are coprime to each other.
By the Chinese Remainder Theorem (CRT), we also obtain the ring isomorphism%\vspace{-1mm}
\begin{equation*}%\label{eq:ring-isom-CRT2}
R_q=\mathbb{F}_q[x]/\langle x^n+1\rangle \cong \bigoplus_{j=0}^{2^\ell -1} \mathbb{F}_q[x]/\langle x^{n/2^\ell} - \zeta_{2^{\ell+1}}^{2j+1}\rangle,
\end{equation*}
where the ring isomorphism is explicitly given by%\vspace{-1mm}
\begin{equation*}%\label{eq:ring-isom-CRT-phi2}
    \mathrm{NTT}_{(\ell,k)} : f \mapsto (f \bmod{x^{n/2^\ell}-\zeta_{2^{\ell+1}}^{2j+1}})_{j=0}^{2^\ell-1},
\end{equation*}
where we identify an element of $R_q$ with its canonical representative of degree less than $n$.

Then, the procedures of incomplete NTT are summarized as follows:

\smallskip

\noindent \textbf{Outline of Incomplete NTT:}
Given $f, g \in R_q$ and an integer $\ell$ with $0 \leq \ell \leq k$, the following procedures compute $f g \in R_q$:
\begin{enumerate}
    \item Compute $\bm{f}\!:=\!\mathrm{NTT}_{(\ell,k)} (f)$ and $\bm{g}\!:=\!\mathrm{NTT}_{(\ell,k)}(g)$, each entry of which is a polynomial of degree less than $n/2^\ell$.
    \item Compute the product $\bm{h}:=\bm{f}\ast \bm{g}$ in the codomain of $\mathrm{NTT}_{(\ell,k)}$ by entry-wise polynomial multiplications with another multiplication method.
    %such as schoolbook, Karatsuba, or Toom--Cook multiplication.
    \item Compute $\mathrm{NTT}_{(\ell,k)}^{-1}(\bm{h})$, and output it.
\end{enumerate}

% \begin{equation*}
%     % h_j \leftarrow g_j+\zeta_{2^{d+1}}^{\frac{e}{2}}g_{j+\frac{m}{2}},\quad
%     % h_{j+\frac{m}{2}} \leftarrow g_j-\zeta_{2^{d+1}}^{\frac{e}{2}}g_{j+\frac{m}{2}}
%         h_j = g_j+\zeta_{2^{t+1}}^{\frac{e}{2}}g_{j+\frac{m}{2}},\quad
%     h_{j+\frac{m}{2}} = g_j-\zeta_{2^{t+1}}^{\frac{e}{2}}g_{j+\frac{m}{2}}
% \end{equation*}
% for each $j=0,\ldots,m/2-1$.
% Similarly, $\Phi_t^{-1}$ is obtained by reversing this decomposition:
% \begin{equation*}
%     g_j = 2^{-1}(h_j+h_{j+\frac{m}{2}}),\quad
%     g_{j+\frac{m}{2}} =  2^{-1} \zeta_{2^{t+1}}^{2^{t}-\frac{e}{2}} (h_{j+\frac{m}{2}}-h_j)
% \end{equation*}
% for each $j=0,\ldots,m/2-1$, where we used $\zeta_{2^{t+1}}^{2^t}=-1$.
% Note that the factor $2^{-1}\in \mathbb{F}_q$ can be precomputed in advance.
% Moreover, the divisions by $2$ arising at each step can be postponed and combined into a single final scaling by $2^{-t}$.

% \smallskip

% The maps $\mathrm{NTT}_{(\ell,k)}$ and $\mathrm{NTT}_{(\ell,k)}^{-1}$ are computed using the same butterfly operations as in the full NTT; see, e.g., \cite{FIPS203,INTT}.
The maps $\mathrm{NTT}_{(\ell,k)}$ and $\mathrm{NTT}_{(\ell,k)}^{-1}$ are computed using the same radix-2 butterfly operations as in standard NTT implementations; see, e.g., \cite{FIPS203,INTT}.
Each level processes $n$ coefficients using $n/2$ multiplications by twiddle factors and $n$ additions/subtractions, yielding a cost of $(\ell n/2)(2\mathsf A+\mathsf M)$ for each transform.
% In the inverse transform, the divisions by $2$ can be postponed and combined into a single multiplication by $2^{-\ell}\in\mathbb{F}_q$, which results in an additional $(n\,\mathbf 1_{\ell>0})\mathsf M$.
In the inverse transform, the divisions by $2$ can be postponed and combined into a single multiplication by $2^{-\ell}\in\mathbb F_q$.
This results in an additional $(n\,\mathbf 1_{\ell>0})\mathsf M$, where $\mathbf 1_{\ell>0}=1$ if $\ell>0$ and $0$ otherwise.
Hence we have the following theorem:

% \smallskip

\begin{theorem}[cf.\ \cite{INTT}, {Observations 3 \& 4}]\label{thm:complexity-INTT}
    Let $C(n;\ell)$ denote the operation cost of multiplying two elements in one CRT component $\mathbb{F}_q[x]/\langle x^{n/2^\ell}-\zeta_{2^{\ell+1}}^{2j+1}\rangle$, which is independent of $j$ in the operation-count model used here.
    %with $d=n/2^t$ and $\zeta = \zeta_{2^{t+1}}$.
    Suppose that $2^{-\ell}\in \mathbb{F}_q$ and the powers of $\zeta_{2^{\ell+1}}$ are computed in advance.
    % Then, the costs of step 1), 2), and 3) for Incomplete NTT are ${\color{red}t n} (2\mathsf{A}+\mathsf{M})$, $2^t C(n;t)$, and $(tn/2) (2\mathsf{A}+\mathsf{M})) + (n\,1_{t>0})\mathsf{M}$, so 
    Then, the total operation cost of incomplete NTT is
    \begin{equation}\label{eq:costINTT}
            3(\ell n/2) (2\mathsf{A} +\mathsf{M}) +   2^\ell C(n;\ell) + (n\,\mathbf{1}_{\ell>0}) \mathsf{M}.
    \end{equation}
    %where $\mathbf{1}_{\ell>0}=1$ if $\ell>0$ and $0$ otherwise.
\end{theorem}

\smallskip

In practice, incomplete NTT is used with an intermediate depth $\ell$ and combined with another multiplication method such as Karatsuba~\cite{Karatsuba} or Toom--Cook~\cite{Toom,Cook} for the smaller subproblems.
For example, Kyber (ML-KEM)~\cite{Kyber,FIPS203} uses $(n,q)=(256,3329)$ with depth $\ell=7$, whereas Dilithium (ML-DSA)~\cite{Dilithium,FIPS204} uses $(n,q)=(256,8380417)$ with depth $\ell=8$ supporting the full NTT.

% Incomplete NTT provides a flexible framework interpolating between no transform and full NTT, and is particularly suitable for hybridization with recursive multiplication methods such as Karatsuba and Toom-Cook~\cite{Toom,Cook}.
% which will be analyzed in detail in Section \ref{sec:main}.

%=======================================
\subsection{Toom--Cook polynomial multiplication}\label{subsec:TC}
%=======================================

We describe Toom--Cook polynomial multiplication~\cite{Toom,Cook} with a splitting parameter $s$ (Toom-$s$ method).
Let $s,d>1$, let $\{ \alpha_0,\alpha_1,\ldots,\alpha_{2s-2}\}$ be a set of $2s-1$ distinct points in $K\cup \{\infty \}$, and let $f,g\in K[x]$ be polynomials of degree less than $d$.
If necessary, replace $d$ by the smallest integer $d'\ge d$ such that $s\mid d'$, and still denote $d'$ by $d$.

\smallskip

\noindent \textbf{Splitting:}
Write $f=\sum_{i=0}^{s-1} f_i x^{id/s}$ and $g=\sum_{i=0}^{s-1} g_i x^{id/s}$ with $f_i,g_i\in K[x]$ and $\deg (f_i),\deg (g_i)<d/s$, and write them as
$F(X)=\sum_{i=0}^{s-1} f_iX^i$ and $G(X)=\sum_{i=0}^{s-1} g_iX^i$ in $K[x][X]$ by putting $X = x^{d/s}$.
% Split $f$ and $g$ into $\ell$ blocks of size $d/\ell$, and regard them as polynomials $F(X),G(X)\in K[x][X]$ with $X=x^{d/\ell}$.
% Let $f(x),g(x) \in K[x]$ be input polynomials of degree less than $d$.
%Split the input polynomials $f(x)$ and $g(x)$ into $\ell$ parts as 
% Write $f = \sum_{i=0}^{\ell-1}f_i(x)x^{id/\ell}$ and $g = \sum_{i=0}^{\ell-1}g_i(x)x^{id/\ell}$, where $\deg(f_i),\deg(g_i)<d/\ell$.
% Putting $X = x^{d/\ell}$, we also write $F(X)=\sum_{i=0}^{\ell-1}f_i(x)X^i$ and $G(X)=\sum_{i=0}^{\ell-1}g_i(x)X^i$.
% We regard $X$ as a variable that is independent of $x$, so that $F(X),G(X)\in K[x][X]$.

\smallskip

\noindent \textbf{Evaluation:}
For each $0 \leq i \leq 2s-2$, compute $F(\alpha_i)$ and $G(\alpha_i)$, which are polynomials in $x$ of degree less than $d/s$.

\smallskip

\noindent \textbf{Recursive multiplication:}
Compute recursively the products $M_i=F(\alpha_i)G(\alpha_i)\in K[x]$ for $0 \leq i \leq 2s-2$.
%, where $F(\alpha_i)$ and $G(\alpha_i)$ are polynomials in $x$ of degree less than $d/\ell$.

\smallskip

\noindent \textbf{Interpolation:}
Recover $H(X) \in K[x][X]$ of degree at most $2s-2$ such that $H(\alpha_i)=M_i$ for $0\leq i \leq 2s-2$.

\smallskip

\noindent \textbf{Re-composition:}
%Substituting $x^{d/\ell}$ to $X$ in $H(X)$, 
We obtain $f(x)g(x)=H(x^{d/s})$.

\smallskip

The case $s=2$ yields Karatsuba multiplication~\cite{Karatsuba}.
When working over $\mathbb F_q[x]/\langle x^d-\omega\rangle$ with $\omega\in \mathbb{F}_q\smallsetminus \{0\}$, one finally reduces modulo $x^d-\omega$.
This reduction costs $(d-1)(\mathsf{A}+\mathsf{M})$, which is reduced to $(d-1)\mathsf{A}$ if $\omega = \pm1$.
% We focus on precise cost estimates for $\ell=4$ and its hybridizations.
% In this paper, however, we focus on more precise cost estimates including constant factors, especially for the case $\ell=4$ (Toom-$4$) and its hybridizations.
It is well-known that the asymptotic complexity of Toom-$s$ is $O(d^{\log_s(2s-1)})$.
Since our target applications involve residue rings defined by $2$-power cyclotomic polynomials and their hybridization with NTT, the degrees $d$ of the resulting subproblems are also powers of $2$.
Therefore, Toom-$4$ is more compatible with recursive splitting than Toom-$3$, since the latter requires padding at multiple recursion levels in power-of-two settings.
% Since our target applications are residue rings defined by $2$-power cyclotomic polynomials and their hybridization with NTT, the relevant subproblem sizes $d$ are $2$-powers.
% Therefore, Toom-$4$ is more compatible with the recursive splitting than Toom-$3$, which would require padding at several levels.
In this paper, we focus on precise cost estimates including constant factors, especially for the case $s=4$ (Toom-$4$) and its hybridizations.

%=====================================
\section{Main results}\label{sec:main}
%=====================================

In this section, we first derive explicit operation counts, including constant factors, for our concrete Toom-$4$ implementation.
% We then analyze two hybrid multiplication methods over the residue ring $R_q$ based on this analysis, and estimate their costs for practical cryptographic parameters.
We then analyze its hybridization with Karatsuba multiplication and incomplete NTT over the residue ring $R_q$.

%====================================================
% \subsection{Detailed complexity analysis for Toom-4}
\subsection{Explicit Toom-4 cost model for incomplete NTT}
%====================================================

Here, we present a concrete Toom-$4$ implementation together with an explicit complexity analysis including constant factors.
For finite-field polynomial multiplication inside CRT components of incomplete NTT, we require operation counts compatible with the cost model of~\cite{INTT}.
% We therefore adopt integer evaluation points, avoid fractional constants, and implement the interpolation scalar multiplications via addition chains, postponing all divisions to a final scaling step.
% Here, we present a concrete implementation of Toom-$4$ together with a detailed complexity analysis including constant factors.
% For our specific setting, namely finite-field polynomial multiplication inside CRT components of incomplete NTT, we require explicit operation counts compatible with the cost model of \cite{INTT}.
The choice of evaluation points and the evaluation/interpolation schedules significantly affects the constant factors of Toom--Cook multiplication; see, e.g., \cite{BZ2007}.
Our purpose here is not to optimize Toom--Cook implementations in full generality, but to obtain a field-operation count suitable for hybridization with incomplete NTT.
For this reason, we adopt integer evaluation points, avoid fractional constants, and implement the scalar multiplications in interpolation via addition chains, postponing all divisions to a final scaling step, thereby yielding a simple and explicit operation count.

Assume that the characteristic of $K$ is neither $2$, $3$, nor $5$.
%As in \cite{BZ2007}, 
% We take $2\ell-1=7$ distinct points to be $\{0,\pm 1, \pm 2, 3,\infty\}$.
We use the seven evaluation points $\{0,\pm1,\pm2,3,\infty\}$.
We also assume that the input size $d$ is divisible by $4$;
otherwise, zero-padding is applied as described in Section~\ref{subsec:TC}.

\smallskip

\noindent \textbf{Splitting:} We split the input polynomials $f(x)$ and $g(x)$ in $K[x]$ of degree less than $d$ as described in Section \ref{subsec:TC}, say
\begin{eqnarray*}
    f(x) &=& f_0 (x) + f_1(x) x^{d/4} + f_2(x) x^{2d/4} + f_3(x) x^{3d/4},\\
    g(x) &=& g_0 (x) + g_1(x) x^{d/4} + g_2(x) x^{2d/4} + g_3(x) x^{3d/4},
\end{eqnarray*}
where $f_i$ and $g_i$ are polynomials of degree less than $d/4$.
Put 
\begin{align*}
    F(X):= f_0 + f_1 X + f_2 X^2 + f_3 X^3,\qquad G(X):= g_0 + g_1 X + g_2 X^2 + g_3X^3.
\end{align*}

\noindent \textbf{Evaluation:} We compute
\begin{align*}
    F(0)=f_0,\qquad F(\infty) = f_3,\qquad F(3) = f_0 + 3f_1 + 9f_2 + 27f_3\\
    F(\pm 1) = f_0\pm f_1+f_2\pm f_3,\qquad F(\pm 2) = f_0 \pm 2f_1 + 4f_2 \pm 8 f_3
\end{align*}
and similarly for $G(0),G(\pm1),G(\pm2),G(3),G(\infty)$, where $F(\infty)$ and $G(\infty)$ denote the leading coefficients of $F(X)$ and $G(X)$, respectively, as polynomials in $X$.
The values $F(\pm1)$, $F(\pm2)$, and $F(3)$ can be computed as follows:
\begin{enumerate}
    \item $a_1:=f_0+f_2$; $a_2:=f_1+f_3$; $a_{3,\pm}:=a_1\pm a_2$ ($=F(\pm 1)$);
    \item $a_4:=f_3 + f_3$ ($=2f_3$); $a_5 := a_4+f_3$ ($=3f_3$);
    \item $a_{6,\pm}:=a_{3,\pm}\pm a_5$ ($=f_0 \pm f_1 + f_2 \pm 4f_3$);
    \item $a_{7,\pm}:=a_{6,\pm}+ f_2$ ($=f_0 \pm f_1 + 2f_2 \pm 4f_3$);
    \item $a_{8,\pm}:=2 a_{7,\pm}$ ($= 2f_0 \pm 2 f_1 + 4 f_2 \pm 8 f_3$);
    \item $a_{9,\pm}:= a_{8,\pm}- f_0$ ($= F(\pm 2)$);
    \item $a_{10}:=2 a_5$ ($=6 f_3$);
    \item $a_{11}:= a_{9,+} - f_0 + a_{10}$ ($=2f_1 + 4 f_2 + 14 f_3$);
    \item $a_{12}:=2 a_{11}$ ($=4 f_1 + 8 f_2 + 28 f_3$);
    \item $a_{13}:= a_{12} + a_{3,-}$ ($=F(3)$).
\end{enumerate}%14 times
% Similarly to the procedures from 2) to 6) for $F(1/2)$, we can compute $F(2)$.
Applying the same procedure to $G$, the evaluation step requires
%$19\times2=38$ additions 
%The same procedure can be applied to compute $G(0)$, $G(\pm1)$, $G(\pm2)$, $G(3)$, and $G(\infty)$.
%Thus, the evaluation step requires 
$19\times 2=38$ additions/subtractions of polynomials of degree less than $d/4$, that is, $38(d/4)$ additions/subtractions in $K$.
% Summing the additions in the 10 steps yields $19$ additions in $K$ for $F$, and the same cost for $G$, resulting in $38$ additions in $K$ in total.

\smallskip

\noindent \textbf{Recursive multiplication:} Compute $M_{\alpha}\!:=\!F(\alpha)G(\alpha)$ for $\alpha\!\in\!\{0,\pm1,\pm2,3,\infty\}$, where $H(X)\!:=\!F(X)G(X)\!\in\! K[x][X]$.
% Compute $H(\alpha)\!=\!F(\alpha)G(\alpha)$ for $\alpha \!\in\! \{ 0,\pm 1, \pm 2, 3,\infty \}$, where $H(X)\!:=\!F(X)G(X)\!\in\! K[x][X]$.
%, and $64\, H(\pm 1/2) = 8\, F(\pm 1/2) \cdot 8\, G(\pm 1/2)$.

\smallskip

\noindent \textbf{Interpolation:}
% Putting $H(X) = \sum_{k=0}^6 h_k X^k$ with $h_k\in K[x]$, we obtain the following system of linear equations:
% \[
% \begin{pmatrix}
%     1 & 0 & 0 & 0 & 0 & 0 & 0\\
%     1 & 1 & 1 & 1 & 1 & 1 & 1\\
%     1 & -1 & 1 & -1 & 1 & -1 & 1 \\ 
%     1 & 2 & 4 & 8 & 16 & 32 & 64\\
%     1 & -2 & 4 & -8 & 16 & -32 & 64\\
%     1 & 3 & 9 & 27 & 81 & 243 & 729\\
%     0 & 0 & 0 & 0 & 0 & 0 & 1
% \end{pmatrix}
% \begin{pmatrix}
%     h_0 \\ h_1\\ h_2\\ h_3\\ h_4\\ h_5\\ h_6
% \end{pmatrix}
% =
% \begin{pmatrix}
%     H(0) \\ H(1)\\ H(-1)\\ H(2)\\ H(-2)\\ H(3)\\ H(\infty)
% \end{pmatrix}
% \]
% Writing $H(X)=\sum_{k=0}^{6} h_k X^k$ with $h_k \in K[x]$, the conditions $H(\alpha)=M_{\alpha}$ for $\alpha\in\{0,\pm1,\pm2,3,\infty\}$ yield a linear system in $h_0,\ldots,h_6$.
% Its coefficient matrix is the corresponding Vandermonde matrix.
Writing $H(X)=\sum_{i=0}^{6} h_i X^i$ for $h_i\in K[x]$ with $\deg(h_i)<2d/4$, the conditions $H(\alpha)=M_{\alpha}$ for $\alpha\in\{0,\pm1,\pm2,3\}$ together with $H(\infty)=h_6=M_{\infty}$ yield a linear system in $h_0,\ldots,h_6$.
Its coefficient matrix is a slightly modified Vandermonde matrix, whose rows correspond to the evaluations at $\alpha=0,1,-1,2,-2,3,\infty$ in this order.
% whose rows correspond to $\alpha=0,1,-1,2,-2,3,\infty$ in this order.
The inverse matrix of the coefficient matrix is
%\vspace{-1mm}
\[
\frac{1}{120}\begin{pmatrix}
    120 & 0 & 0 & 0 & 0 & 0 & 0\\
    -40 & 120 & -60 & -30 & 6 & 4 & -1440\\
    -150 & 80 & 80 & -5 & -5 & 0 & 480\\ 
    50 & -70 & -5 & 35 & -5 & -5 & 1800\\
    30 & -20 & -20 & 5 & 5 & 0 & -600\\
    -10 & 10 & 5 & -5 & -1 & 1 & -360 \\
    0 & 0 & 0 & 0 & 0 & 0 & 120
\end{pmatrix}
\]
% \vspace{-1mm}
% Using this inverse matrix, each $h_i$ is expressed as a linear combination of the values $M_\alpha(=H(\alpha))$.
% For example, 
Using this inverse matrix, each $h_i$ can be expressed as a linear combination of the values $M_\alpha=H(\alpha)$.
For example, %\vspace{-1mm}
\begin{align*}
    h_1
    =120^{-1}\bigl(&-40H(0)+120H(1)-60H(-1)-30H(2)+6H(-2)+4H(3)-1440H(\infty)\bigr),%\\[-6mm]
\end{align*}
and $h_2,\ldots,h_5$ are obtained similarly.

Hence the nontrivial interpolation coefficients have denominator $120$, so we work with the scaled quantities $h_i':=120h_i$ throughout the recursion.
Thus, after $L$ recursive Toom-$4$ levels, the output is scaled by $120^L$ and recovered by a single final multiplication by the precomputed inverse $120^{-L}\in K$.
The recursion is terminated at a suitable base case, where Karatsuba or schoolbook multiplication is used.

For this interpolation step, we first compute several scalar multiples of each $H(\alpha)$ appearing in the above inverse matrix.
% For example, we require $120 H(0)$, $40 H(0)$, $150 H(0)$, $50H(0)$, $30H(0)$, and $10H(0)$ when $\alpha=0$.
% These scalar multiples are computed using {\it addition chains}.
% Namely, an addition chain for a positive integer $m$ is a sequence $1=b_0,b_1,\ldots,b_r=m$ such that each $b_t$ $(t\ge1)$ is of the form $b_i+b_j$ for some $0\le i,j<t$; each step counts as one addition for scalar multiplication in $K[x]$.
% For the interpolation step, we must compute several scalar multiples of each $H(\alpha)$ appearing in $120A^{-1}$.
Thus, rather than generating a single scalar multiple, we consider addition chains for a prescribed set of target multiples.
For a finite set $S$ of positive integers, an {\it addition chain for $S$} is a sequence $1=b_0,b_1,\ldots,b_r$ such that each $b_m$ with $1\leq m\leq r$ is of the form $b_i+b_j$ for some $0\le i,j<m$, and $S\subset\{b_0,\ldots,b_r\}$.
% Its cost is $r$, i.e., the number of additions.
Using such an addition chain, the required scalar multiples can be computed with $r$ additions.
%The boldface entries below indicate the target sets required for interpolation.

% The following chains minimize the number of additions needed to generate all required scalar multiples within the prescribed target sets.
% The optimality of these chains for the prescribed target sets was verified by exhaustive enumeration up to the relevant lengths, together with elementary lower-bound arguments for the final large targets. 
% The verification scripts are included in our public repository.
The following chains minimize the number of additions needed to generate all required scalar multiples within the prescribed target sets.
The optimality of these chains for the prescribed target sets was verified by exhaustive enumeration up to the relevant lengths, together with elementary lower-bound arguments for the remaining large targets. 
% Their optimality was verified by exhaustive enumeration up to the relevant lengths, together with elementary lower-bound arguments for the remaining large targets.
The verification scripts are included in our public repository~\cite{GitHub}.
%\vspace{-1mm}
\begin{align*}
    H(0): & \, 1,\ 2,\ 4,\ 8,\ {\bf 10},\ 20,\ {\bf 30},\ {\bf 40},\ {\bf 50},\ 100,\ {\bf 120},\ {\bf 150};\\
    H(1): & \, 1,\ 2,\ 4,\ 8,\ {\bf 10},\ {\bf 20},\ 30,\ 40,\ {\bf 70},\ {\bf 80},\ {\bf 120} ;\\
    H(-1): & \, 1,\ 2,\ 4,\ {\bf 5},\ {10},\ {\bf 20},\ 40,\ {\bf 60},\ {\bf 80} ;\\
    H(2): & \, 1,\ 2,\ 4,\ {\bf 5},\ {10},\ 20,\ {\bf 30},\ {\bf 35} ;\\
    H(-2): & \, 1,\ 2,\ 4,\ {\bf 5},\ {\bf 6};\\
    H(3): & \, 1,\ 2,\ {\bf 4},\ {\bf 5} ;\\
    H(\infty): & \, 1,\ 2,\ 4,\ 8,\ 10,\ 20,\ 40,\ 80,\ {\bf 120},\ 240,\ {\bf 360},\ {\bf 480},\ {\bf 600},\ 1080,\ {\bf 1440},\ {\bf 1800}.%\\[-7mm]
\end{align*}
%\vspace{-2mm}
%where each non-initial entry is obtained as the sum of two earlier entries.
The boldface entries indicate the scalar multiples appearing in the above inverse matrix.
% Thus, all required scalar multiples are obtained with $11+10+8+7+4+3+15 = 58$ additions of polynomials of degree less than $2d/4$.
% This formulation allows us to control the constant factors explicitly, which is crucial for comparing multiplication strategies in the incomplete NTT framework.
Thus, all required scalar multiples are obtained with $11+10+8+7+4+3+15 = 58$ additions of polynomials of degree less than $2d/4$, namely $58(2d/4)$ additions in $K$.
% This formulation allows us to control the constant factors explicitly, which is crucial for comparing multiplication strategies in the incomplete NTT framework.
% This explicit control of scalar multiplications is essential in the incomplete NTT setting, where constant factors directly affect the optimal hybrid strategy.
This explicit control of scalar multiplications is crucial in the incomplete NTT setting, where constant factors directly affect the optimal hybrid strategy.

Once these scalar multiples are prepared, the scaled coefficients $h_i'$ are recovered by additions/subtractions.
% For instance, the above expression for $h_1'$ requires $6$ additions once the relevant scalar multiples have been computed.
For instance, the scaled coefficient $h_1'$ is obtained with $6$ additions/subtractions once the relevant scalar multiples have been computed.
Similarly, $h_2'$, $h_3'$, $h_4'$, and $h_5'$ are obtained with $5$, $6$, $5$, and $6$ additions/subtractions, respectively.
Hence the polynomials $h_1',\ldots,h_5'$ are recovered with $28$ additions/subtractions of polynomials of degree less than $2d/4$, that is, $28(2d/4)$ additions/subtractions in $K$.

\smallskip

\noindent \textbf{Re-composition:}
At the scaled level, recovering $120H(x^{d/4})$ from $h_0',\ldots,h_6'$ requires $6$ additions on overlapping blocks of size $d/4$, namely $6(d/4)$ additions in $K$.
Indeed, since each $h_i'$ has degree less than $2d/4$, the supports of $h_0',\ h_1'x^{d/4},\ \ldots,\ h_6'x^{6d/4}$ overlap only with adjacent blocks.
Hence the recomposition can be carried out with $6$ additions on overlapping blocks of size $d/4$.

\smallskip

Collecting the above counts, one Toom-$4$ level requires $38(d/4) + (58+28)(2d/4) + 6(d/4)=54d$ additions in $K$, where the three terms correspond to evaluation, interpolation, and re-composition, respectively.
% After the recursion terminates, one final rescaling by the precomputed inverse of the accumulated factor is applied to all coefficients of the output polynomial, which requires $2d-1$ further multiplications in $K$.
The final rescaling by the precomputed inverse of the accumulated factor to the output polynomial requires $2d-1$ further multiplications in $K$.

% \smallskip

\begin{prop}\label{prop:main}
Let $T_{\mathsf A}(d)$ and $T_{\mathsf M}(d)$ denote the numbers of additions/subtractions and multiplications in $K$ required by the recursive part of the proposed Toom-$4$ implementation on inputs of size $d$, excluding the final rescaling.
Then we have $T_{\mathsf A}(d)=7\,T_{\mathsf A}(d/4)+54d$ and $T_{\mathsf M}(d)=7\,T_{\mathsf M}(d/4)$.
In addition, the final rescaling step requires $2d-1$ further multiplications in $K$.
\end{prop}

% \smallskip

% This follows directly from the above construction 
The claim follows directly from the above construction by counting the number of arithmetic operations in each step.

\smallskip

\begin{remark}
% Bodrato and Zanoni~\cite{BZ2007} studied the optimization of Toom--Cook evaluation and interpolation matrices, including choices of evaluation points such as fractional points. 
% In contrast, our implementation avoids fractional evaluation points such as $\pm 1/2$ and postpones divisions to a final scaling step. 
% This design choice simplifies the operation-count model over $\mathbb F_q$ and avoids additional scaling factors, which would otherwise complicate integration into the incomplete NTT cost framework.
% We also tested interpolation matrices arising from fractional evaluation points such as $\pm 1/2$. 
% After clearing denominators and applying the same addition-chain optimization to the resulting scalar multiples, these choices required more additions than the present set of points. 
% Thus, for the present operation-count model, the integer evaluation points $\{0,\pm1,\pm2,3,\infty\}$ give a simpler and cheaper interpolation schedule.
% Bodrato and Zanoni~\cite{BZ2007} studied the optimization of Toom--Cook evaluation and interpolation matrices, including the use of fractional evaluation points such as $\pm 1/2$. 
Prior work on efficient Toom--Cook multiplication~\cite{BZ2007,Bodrato2007} studied optimized evaluation and interpolation schemes, including fractional evaluation points such as $\pm 1/2$.
In contrast, we restrict to integer evaluation points and postpone all divisions to a final scaling step, which simplifies the operation-count model over $\mathbb F_q$ and facilitates integration into the incomplete NTT framework.
We also tested fractional evaluation points;
after clearing denominators and applying the same addition-chain optimization, they required more additions/subtractions than the present choice.
Thus, the set $\{0,\pm1,\pm2,3,\infty\}$ of evaluation points gives a simpler and cheaper interpolation schedule in our setting.
\end{remark}

%===============================================
\subsection{Hybrid multiplication over a residue ring}\label{subsec:hybrid}
%===============================================

% As in Section~\ref{subsec:IncompleteNTT}, let $q$ be an odd prime,
% let $n=2^k$, and assume that $2^{\ell+1}\mid(q-1)$.
% From now on we set $K=\mathbb{F}_q$ and $R_q = \mathbb{F}_q[x]/\langle x^n+1\rangle$.
From now on, let $q$ be an odd prime and set $K=\mathbb F_q$.
As in Section~\ref{subsec:IncompleteNTT}, let $n=2^k$, assume that $2^{\ell+1}\mid(q-1)$, and put $R_q=\mathbb F_q[x]/\langle x^n+1\rangle$.
In this subsection, we instantiate the abstract cost function $C(n;\ell)$ in Theorem~\ref{thm:complexity-INTT} by a concrete hybrid multiplication method over $R_q$.
% Throughout this subsection, let $d$ denote the subproblem size, and we assume that $d$ is a power of $2$.
Recall that $n$ is a power of $2$ in our setting (Section~\ref{subsec:IncompleteNTT}).
Throughout this subsection, let $d$ denote the subproblem size; in particular, $d=n/2^\ell$ is also a power of $2$.

\smallskip

\noindent \textbf{Toom-4/Karatsuba hybrid.}
We apply Toom-$4$ recursively for $L$ levels with Karatsuba multiplication as the base case.
Let $T_{\mathsf M}(d;L)$ and $T_{\mathsf A}(d;L)$ denote the numbers of multiplications and additions/subtractions in $\mathbb F_q$ for multiplying two polynomials of degree less than $d$; the cost of the reduction modulo $x^d-\omega$ is counted separately.
%excluding the reduction modulo $x^d-\omega$.
% By Proposition~\ref{prop:main}, we have
By Proposition~\ref{prop:main} and summing the geometric series of the $54d$-terms, we have
\begin{align*}
T_{\mathsf M}(d;L)&=7^L\,T_{\mathsf M}^{\rm Kar}(d/4^L)+(2d-1)\mathbf 1_{L>0},\\
T_{\mathsf A}(d;L)&=7^L\,T_{\mathsf A}^{\rm Kar}(d/4^L)+72d\bigl((7/4)^L-1\bigr),
\end{align*}
where $\mathbf 1_{L>0}=1$ if $L>0$ and $0$ otherwise.
Here $T_{\mathsf M}^{\rm Kar}(d')$ and $T_{\mathsf A}^{\rm Kar}(d')$ denote the operation counts of Karatsuba multiplication for input size $d'$, estimated by $(d')^{\log_2 3}$ and $8((d')^{\log_2 3}-d')$ respectively.
These counts follow from the standard recursive operation count for Karatsuba multiplication; see, e.g., the proof of Theorem~8.3 in~\cite{vzGG}.
Consequently we obtain
\begin{align*}
T_{\mathsf M}(d;L) &= 7^L(d/4^L)^{\log_2 3} + (2d-1)\mathbf 1_{L>0},\\
T_{\mathsf A}(d;L) &= 7^L\cdot 8\bigl((d/4^L)^{\log_2 3}-d/4^L\bigr)+72d\bigl((7/4)^L-1\bigr).
\end{align*}

% \smallskip

\noindent \textbf{Incomplete NTT with Toom-4/Karatsuba hybrid.}
As in Subsection~\ref{subsec:IncompleteNTT},
let $\ell$ be the NTT depth.
Following~\cite{INTT}, let $\alpha$ and $\mu$ denote the costs of one addition/subtraction and one multiplication in $\mathbb F_q$, respectively.
We normalize by $\mu$ and put $w:=\alpha/\mu$.
For one subproblem of size $d$, define
\begin{equation}\label{eq:CwdL}
C_w(d;L):=T_{\mathsf M}(d;L)+w\,T_{\mathsf A}(d;L).
\end{equation}
Substituting this into the cost formula~\eqref{eq:costINTT}
of Theorem~\ref{thm:complexity-INTT}
with $d=n/2^\ell$,
we obtain the following weighted cost model:
\begin{equation}\label{eq:CwntL}
\begin{aligned}
C_w(n;\ell,L)
&=
2^\ell C_w(n/2^\ell;L)
+
({3\ell n}/2)(1+2w)
+
(2n-2^\ell)\mathbf 1_{\ell>0}
+
w(n-2^\ell),
\end{aligned}
\end{equation}
where the last two terms account for the final scaling and the reduction modulo
$x^{n/2^\ell}-\zeta_{2^{\ell+1}}^{2j+1}$
in each CRT component.

For a fixed triple $(n,\ell,w)$,
only the first term in \eqref{eq:CwntL}
depends on $L$.
Therefore, for each pair $(n,\ell)$,
the optimal recursion depth is obtained by minimizing
\eqref{eq:CwdL} with $d=n/2^\ell$.

\subsection{Parameter optimization and comparison}\label{subsec:para}
%=======================================================

% We evaluate the cost model in Subsection~\ref{subsec:hybrid}.
% To highlight the effect of multiplication algorithms, we fix $w=0.2$ (i.e., $\mu/\alpha=5$ as in \cite{INTT}), which reflects a typical setting where additions are cheaper than multiplications.

We evaluate the cost model in Subsection~\ref{subsec:hybrid}.
To highlight the effect of multiplication algorithms, we fix $w=0.2$ (i.e., $\mu/\alpha=5$ as in \cite{INTT}), which reflects a typical setting where additions/subtractions are cheaper than multiplications.
% Other values of $w$ lead to qualitatively similar behavior, and the case $w=0.2$ is chosen as a representative setting following~\cite{INTT}.

For $w=0.2$, Toom-$4$ becomes effective only for sufficiently large subproblems:
Minimizing $C_w(d;L)$ in \eqref{eq:CwdL} over $L$ shows that the optimal recursion depth is $L=0$ for $d\le 128$, $L=1$ for $d=256$, and $L=2$ for $d=512,1024$.
In particular, for $d\le 128$, the hybrid method reduces to pure Karatsuba.
% Since Karatsuba strictly dominates schoolbook multiplication for all parameter sizes considered here, we omit schoolbook from the comparison.
Since Karatsuba has a lower cost than schoolbook multiplication in our model
for all parameter sizes considered here, we omit schoolbook from the comparison.
% Since Karatsuba (the case $L=0$ in the Toom-$4$/Karatsuba hybrid) already dominates schoolbook multiplication for the parameter range considered here, we omit schoolbook from the comparison below.

We compare three multiplication strategies under the same NTT depth $\ell$:
Karatsuba ($L=0$), pure Toom-$4$, and the proposed Toom-$4$/Karatsuba hybrid.
Here, ``pure Toom-$4$'' means that Toom-$4$ is applied as long as possible and Karatsuba multiplication is used at the final base level.
The maximal recursion depth is denoted by $L_{\max}$, while $L_{\rm opt}$ denotes the value minimizing $C_w(d;L)$ in \eqref{eq:CwdL}.
The results for $n\in \{256,512,1024\}$ are shown in Table~\ref{tab:compare}.

\begin{table}[t]
\centering
\caption{
Weighted cost $C_{w}(n;\ell,L)$ with $w=0.2$ and $n\in \{256,512,1024\}$ for multiplication over $\mathbb{F}_q[x]/\langle x^n+1\rangle$.
Here $\ell$ is the NTT depth, $d=n/2^\ell$ is the subproblem size, and $w=0.2$ is the relative cost of additions/subtractions.
The columns correspond to Karatsuba ($L=0$), pure Toom-$4$ ($L=L_{\max}$),
and the hybrid method ($L=L_{\rm opt}$).
Costs are measured in units of one multiplication in $\mathbb{F}_q$.
}
\label{tab:compare}
\vspace{2mm}
\begin{tabular}{c|c|c|rrr}
\hline
% \multirow{2}{*}{$n$} & \multirow{2}{*}{$t$} & \multirow{2}{*}{$d$} & \multicolumn{1}{c}{Karatsuba} & \multicolumn{1}{c}{Toom-$4$} & \multicolumn{1}{c}{Hybrid} \\
%  & & & \multicolumn{1}{c}{($L=0$)} & \multicolumn{1}{c}{($L=L_{\rm max}$)} & \multicolumn{1}{c}{($L=L_{\rm opt}$)} \\
$n$ & $\ell$ & $d$ & \multicolumn{1}{c}{Karatsuba} & \multicolumn{1}{c}{Toom-$4$} & \multicolumn{1}{c}{Hybrid} \\
\hline

\multirow{2}{*}{256}
 & 0 & 256 & 16700.0 & 33851.0 & 15877.8 ($L_{\rm opt}=1$) \\
 & 1 & 128 & 12061.2 & 20834.4 & 12061.2 ($L_{\rm opt}=0$) \\
 % & 3 & 64 & 17611 & 26702 & 17611 ($L=0$) \\
\hline
\multirow{3}{*}{512}
 & 0 & 512 & 50458.8 & 73945.8 & 44781.0 ($L_{\rm opt}=2$) \\
 & 1 & 256 & 35497.2 & 69799.2 & 33852.8 ($L_{\rm opt}=1$) \\
 & 2 & 128 & 25197.6 & 42744.0 & 25197.6 ($L_{\rm opt}=0$) \\
 % & 3 & 64 & 17611 & 26702 & 17611 ($L=0$) \\
\hline
\multirow{4}{*}{1024}
 & 0 & 1024 & 152093.6 & 246333.8 & 120521.4 ($L_{\rm opt}=2$) \\
 & 1 & 512 & 105114.0 & 152088.0 & 93758.4 ($L_{\rm opt}=2$) \\
 & 2 & 256 & 73144.8 & 141748.8 & 69856.0 ($L_{\rm opt}=1$) \\
 & 3 & 128 & 52545.6 & 87638.4 & 52545.6 ($L_{\rm opt}=0$) \\
\hline
\end{tabular}
\vspace{-4mm}
\end{table}

% n = 512
% t=0, d=512, Kar=51481.8, Toom4(Lmax=4)=74968.8, Hyb(L=2)=45804.0
% t=1, d=256, Kar=35138.8, Toom4(Lmax=4)=69440.8, Hyb(L=1)=33494.4
% t=2, d=128, Kar=24480.8, Toom4(Lmax=3)=42027.2, Hyb(L=0)=24480.8
% t=3, d=64, Kar=17611.2, Toom4(Lmax=3)=39168.0, Hyb(L=0)=17611.2
% n = 1024
% t=0, d=1024, Kar=154140.6, Toom4(Lmax=5)=248380.8, Hyb(L=2)=122568.4
% t=1, d=512, Kar=104397.2, Toom4(Lmax=4)=151371.2, Hyb(L=2)=93041.6
% t=2, d=256, Kar=71711.2, Toom4(Lmax=4)=140315.2, Hyb(L=1)=68422.4
% t=3, d=128, Kar=50395.2, Toom4(Lmax=3)=85488.0, Hyb(L=0)=50395.2
% t=4, d=64, Kar=36656.0, Toom4(Lmax=3)=79769.6, Hyb(L=0)=36656.0

% \smallskip

Table~\ref{tab:compare} shows that, for fixed $\ell$, the proposed hybrid method consistently improves on pure Toom-$4$ and, for sufficiently small $\ell$, also improves on Karatsuba.
This suggests that, beyond previous analyses focusing mainly on schoolbook or Karatsuba multiplication, Toom-$4$ can provide additional gains in the limited-depth regime.
As $\ell$ increases, the subproblem size $d=n/2^\ell$ becomes smaller, and the optimal choice eventually becomes $L=0$, so that the hybrid method coincides with Karatsuba.

% Thus, for parameters admitting large NTT depths, such as Kyber, the improvement by Toom-$4$ is limited.
Thus, for parameter sets with large admissible NTT depths, such as the recommended parameters of Kyber, the improvement by Toom-$4$ is limited.
In contrast, when the admissible depth is restricted by the $2$-adic valuation of $q-1$, the proposed hybrid yields a noticeable reduction in cost.

%=====================================
\section{Experimental validation}
%=====================================

We provide an experimental validation of the proposed hybrid method.
The goal is not to obtain precise quantitative agreement with the theoretical cost model, but to confirm the qualitative behavior predicted in Section~\ref{subsec:para}.
All experiments were conducted on a standard laptop environment (Python 3, single-threaded); see~\cite{GitHub} for details.
%We implemented Karatsuba, pure Toom-$4$, and the proposed hybrid method.
We implemented incomplete NTT together with Karatsuba, pure Toom-$4$, and the proposed hybrid method.
Polynomial multiplication is performed over $\mathbb F_q[x]/\langle x^n+1\rangle$ with $n\in\{256,512,1024\}$.
To make the limited-depth regime visible, we choose primes $q$ of comparable size
but with different $2$-adic valuations of $q-1$, which directly control the maximal admissible depth of incomplete NTT.
For a prime $q$, the maximal admissible incomplete NTT depth is
$\ell_{\max} = \min\{v_2(q-1)-1,\ \log_2 n\}$, where $v_2(q-1)$ denotes the $2$-adic valuation of $q-1$.
While the case of Kyber was experimentally studied in \cite{INTT}, we choose NTT-unfriendly prime moduli (primes for which the admissible NTT depth is limited) of comparable size near the Dilithium modulus:
$q=8380403,8380381,8380249,8380369$, for which $\ell_{\max}=0,1,2,3$, respectively.
For each $(n,q)$, we apply incomplete NTT with depth $\ell=\ell_{\max}$, and report the median running time over 100 random inputs.

\begin{table}[t]
\centering
\caption{
Running time (ms) for multiplication over $\mathbb F_q[x]/\langle x^n+1\rangle$.
Here $\ell_{\max} = \min\{v_2(q-1)-1,\ \log_2 n\}$ is the maximal admissible NTT depth, and $d=n/2^{\ell_{\max}}$ is the subproblem size.
For incomplete NTT with depth $\ell=\ell_{\max}$, we compare Karatsuba ($L=0$),
pure Toom-$4$, and the hybrid method ($L=L_{\mathrm{opt}}$).
Each value is the median over 100 random inputs.
}
\label{tab:exp}
\vspace{2mm}
\begin{tabular}{c|c|c|c|rrr}
\hline
$n$ & $q$ & $\ell_{\max}$ & $d$ & Karatsuba & Toom-$4$ & Hybrid \\
% \hline
% \multirow{2}{*}{$n$} & \multirow{2}{*}{$q$} & \multirow{2}{*}{$t_{\max}$} & \multirow{2}{*}{$d$} & \multicolumn{1}{c}{Karatsuba} & \multicolumn{1}{c}{Toom-$4$} & \multicolumn{1}{c}{Hybrid} \\
% & & & & \multicolumn{1}{c}{($L=0$)} & \multicolumn{1}{c}{($L=L_{\rm max}$)} & \multicolumn{1}{c}{($L=L_{\rm opt}$)} \\
\hline
% \multirow{2}{*}{256}
%  & 8380403 & 0 & 256 & 17.250 & 30.973 & 14.487 \\
%  & 8380381 & 1 & 128 & 12.030 & 15.952 & 12.030 \\
% % 256 & 8380249 & 2 & 32  & ... & ... & ... \\
% \hline
% \multirow{3}{*}{512}
%  & 8380403 & 0 & 512 & 106.193 & 112.972 & 77.214 \\
%  & 8380381 & 1 & 256 & 64.657 & 114.045 & 54.560 \\
%  & 8380249 & 2 & 128 & 27.655 & 36.177 & 27.655 \\
% % 512 & 8380369 & 3 & 32  & ... & ... & ... \\
% \hline
% \multirow{4}{*}{1024}
% & 8380403 & 0 & 1024 & 234.723 & 308.520 & 156.108 \\
% & 8380381 & 1 & 512 & 152.812 & 171.839 & 113.301 \\
% & 8380249 & 2 & 256 & 144.147 & 251.387 & 122.747 \\
% & 8380369 & 3 & 128 & 105.562 & 135.290 & 105.562 \\

\multirow{2}{*}{256} & 8380403 & 0 & 256 & 16.722 & 32.873 & 14.182 \\
& 8380381 & 1 & 128 & 11.130 & 15.405 & 11.130 \\ \hline
\multirow{3}{*}{512} & 8380403 & 0 & 512 & 50.738 & 57.431 & 36.901 \\
& 8380381 & 1 & 256 & 33.975 & 66.562 & 28.998 \\
& 8380249 & 2 & 128 & 22.812 & 31.415 & 22.812 \\ \hline
\multirow{4}{*}{1024} & 8380403 & 0 & 1024 & 152.838 & 237.411 & 104.767 \\
& 8380381 & 1 & 512 & 101.967 & 114.807 & 74.587 \\
& 8380249 & 2 & 256 & 68.567 & 133.386 & 58.385 \\
& 8380369 & 3 & 128 & 45.749 & 63.358 & 45.749 \\

\if0
\multirow{2}{*}{256} & 8380403 & 0 & 256 & 18.038 & 35.723 & 15.205 \\
 & 8380381 & 1 & 128 & 14.101 & 20.804 & 14.101 \\ \hline
\multirow{3}{*}{512} & 8380403 & 0 & 512 & 100.673 & 116.393 & 72.920 \\
 & 8380381 & 1 & 256 & 71.378 & 140.483 & 60.461 \\
 & 8380249 & 2 & 128 & 38.173 & 52.797 & 38.173 \\ \hline
\multirow{4}{*}{1024} & 8380403 & 0 & 1024 & 409.078 & 676.963 & 282.469 \\
& 8380381 & 1 & 512 & 250.545 & 284.400 & 183.437 \\
& 8380249 & 2 & 256 & 194.363 & 384.626 & 158.293 \\
& 8380369 & 3 & 128 & 171.974 & 233.912 & 171.974 \\
\fi

\hline
\end{tabular}
\vspace{-3mm}
\end{table}

Table~\ref{tab:exp} reports the running times.
For each instance, the hybrid method uses the optimal recursion depth $L_{\mathrm{opt}}$ for the corresponding subproblem size.
Note that when $L_{\mathrm{opt}}=0$, the hybrid method reduces to Karatsuba; this explains the identical entries in the corresponding rows.
The results are consistent with the theoretical prediction.
When the admissible NTT depth is very limited (e.g., $\ell_{\max}=0$), the hybrid method significantly outperforms Karatsuba.
As the admissible depth increases, the advantage gradually diminishes, and eventually disappears when the subproblem size becomes too small.
This supports the conclusion that the proposed method is mainly beneficial in the limited-depth regime of incomplete NTT.

The implementation used for the experiments is publicly available at~\cite{GitHub}.

%===================================
\section{Conclusion and future work}
%===================================

In this paper, we presented a detailed complexity analysis of Toom-$4$ polynomial multiplication including constant factors, and integrated it into the incomplete NTT framework over $R_q=\mathbb{F}_q[x]/\langle x^n+1\rangle$.
Based on this analysis, we derived explicit cost formulas and identified the optimal recursion depth in a weighted cost model.
Our current implementation is in Python; however, the proposed cost model is independent of implementation details and is intended to guide optimized low-level implementations.

Our results show that Toom-$4$ provides additional gains mainly in the limited-depth regime of incomplete NTT, where the admissible depth is restricted by the $2$-adic valuation of $q-1$.
In particular, the hybrid method is most effective when the transform depth is minimal, highlighting a practically relevant regime where Toom-$4$ becomes advantageous.

Future work includes optimized low-level implementations (e.g., in C or C++) and further evaluation in concrete lattice-based cryptographic schemes.

\section*{Acknowledgment}
The authors thank Masaya Yasuda for helpful comments.
Preliminary results of this work were presented at the poster session of the 2026 Symposium on Cryptography and Information Security (SCIS 2026), a Japanese conference on cryptography and information security, where the authors received valuable comments and suggestions, particularly on hybridization with NTT.
This work was partially supported by Comprehensive Research Organization, Fukuoka Institute of Technology.
This work was also partially supported by JST CREST Grant Number JPMJCR2113 and JSPS KAKENHI Grant Number 25K00140.

\end{document}